\documentclass[10pt,conference]{IEEEtran}
\usepackage[utf8]{inputenc}
\usepackage{cite}
\usepackage{amsmath,amssymb,amsfonts}
\usepackage{algorithmic}
\usepackage{graphicx}
\usepackage{textcomp}
\usepackage{xcolor}
\usepackage{hyperref}
\usepackage{url}
\usepackage{graphicx}
\usepackage{array}
\newcolumntype{P}[1]{>{\centering\arraybackslash}p{#1}}
\usepackage{booktabs}

\graphicspath{{figures}}

\makeatletter
\def\ps@IEEEtitlepagestyle{
	\def\@oddfoot{\mycopyrightnotice}
	\def\@evenfoot{}
}
\def\mycopyrightnotice{
	{\footnotesize
		\begin{minipage}{\textwidth}
			\centering
			\textcopyright~2023 IEEE.  Personal use of this material is permitted.  Permission from IEEE must be obtained for all other uses, in any current or future media, including reprinting/republishing this material for advertising or promotional purposes, creating new collective works, for resale or redistribution to servers or lists, or reuse of any copyrighted component of this work in other works.
		\end{minipage}
	}
}

\begin{document}

\title{On the Need for Artifacts to Support Research on Self-Adaptation Mature for Industrial Adoption}

\author{\IEEEauthorblockN{Danny Weyns}
\IEEEauthorblockA{
\textit{Katholieke Universiteit Leuven, Belgium}\\
\textit{Linnaeus University Sweden}\\
danny.weyns@kuleuven.be}\vspace{-10pt}
\and
\IEEEauthorblockN{Thomas Vogel}
\IEEEauthorblockA{
\textit{Humboldt-Universit\"{a}t} zu Berlin\\
Berlin, Germany \\
thomas.vogel@informatik.hu-berlin.de}\vspace{-10pt}
}

\maketitle

\begin{abstract}

Despite the vast body of knowledge developed by the self-adaptive systems community and the wide use of self-adaptation in industry, it is unclear whether or to what extent industry leverages output of academics. Hence, it is important for the research community to answer the question: \textit{Are the solutions developed by the self-adaptive systems community mature enough for industrial adoption?} Leveraging a set of empirically-grounded guidelines for industry-relevant artifacts in self-adaptation, we develop a position to answer this question from the angle of using artifacts for evaluating research results in self-adaptation, which is actively stimulated and applied by the community.  

\end{abstract}

\def\arraystretch{1.25}

\section{Introduction}

Over the past two decades, self-adaptation has become an established field of research. 
Since pioneering efforts of Oreizy et al.~\cite{oreizy1999aba}, Kephart and Chess~\cite{kephart2003vision}, and Garlan et al.~\cite{garlan2004rainbow}, the field has matured and generated a large body of knowledge~\cite{weyns2021introduction}. 

A recent large-scale survey~\cite{3524844.3528077,survey-report} provided evidence that the principles of self-adaptation are also widely applied in industry. Among 184 participants working in a wide variety of domains, about 54\% expressed to have worked with concrete self-adaptive systems. 
The main reported benefits of applying self-adaptation are improved utility (in robustness and performance), savings (costs and resources), and improved human interaction (user experience and engineers support). Yet, a majority of participants face difficulties when engineering or maintaining self-adaptive systems, mainly with reliable/optimal design, design complexity, and tuning/debugging. About half of the practitioners report that they would appreciate support from researchers to deal with problems they face. 

Despite the vast body of knowledge developed by academics and the wide use of self-adaptation in industry, it is unclear to what extent research output is used in industry. Insights into this are important as it will help researchers to position their efforts with respect to industrial needs and make well-informed decisions to set future research objectives. It will also benefit practitioners directing them towards relevant sources of information and identifying opportunities for collaboration with researchers to address the problems they face. Hence, it is important for the community to answer the question: 
\vspace{5pt}\\
    \mbox{\ \ \ }{\textit{Are the solutions developed by the self-adaptive systems \newline \mbox{\ \ \ }community mature enough for industrial adoption?} 
\vspace{5pt}\\
In this paper, we develop a position to answer this question from the angle of how research results in the self-adaptive systems community are evaluated. More specifically, we focus on the use of artifacts for evaluating and comparing new research results in self-adaptation, which is actively stimulated by the community. With artifact, we refer here to a tangible object that is created with the intention to drive research advances and compare and contrast new and alternative approaches.\footnote{\url{https://conf.researchr.org/track/seams-2022/seams-2022-papers\#Artifacts}}

\small
\begin{table*}[!t]
\renewcommand{\arraystretch}{1.08}
\begin{center}
\caption{Industry-relevance of a selection of artifacts for self-adaptation using the guidelines of~\cite{10.1145/3561846.3561852}. \newline (Px: required guideline, Rx: recommended, Dx: desirable) \vspace{-5pt}}
\label{tab:overview}
\begin{tabular}{p{5.4cm}P{1.2cm}P{1cm}P{1.5cm}P{1.5cm}P{1.1cm}P{1.5cm}P{1.3cm}} 
\toprule 
\textbf{Guidelines} & PLEGOs~\cite{9462025} & BSN~\cite{9462020} & RoboMAX~\cite{RoboMax} & RDMSim~\cite{9462042} & EWS~\cite{9799868} & SEAByTE~\cite{9799807} & Simdex~\cite{9800016} \\ \midrule 
\textbf{P1.1: }Consider management of real-world data & 0 & 0 & 0 & 0 & 0 & 0 & $+$ \\ 
\textbf{P1.2: }Take into account humans-on-the-loop & 0 & 0 & 0 & 0 & 0 & $+$ & 0   \\  
\textbf{P1.3: }Consider industrial adoption of solutions & 0 & 0 & 0 & 0 & 0 & $++$ & 0  \\ 
\textbf{P2: }Take into account industry scale  & 0 & 0 & 0 & 0 & 0 & 0 & 0  \\ 
\textbf{P3: }Offer industry-relevant metrics  & $+$ & $+$ & $+$ & $+$ & $+$ & $+$ & $+$  \\ 
\textbf{P4: }Use an open policy  & $++$ & $++$ & $++$ & $++$ & $++$ & $++$ &  $++$ \\ \midrule 
\textbf{R1: }Integrate with standard technologies  & $+$ & $++$ & 0 & $+$ & $+$ & $++$ & $+$   \\ 
\textbf{R2: }Process-related artifacts  & 0 & 0 & 0 & 0 & 0 & 0 & 0  \\ 
\textbf{R3: }Generality and extensibility  & 0 & 0 & $+$ & 0 & 0 & 0 & 0  \\ 
\textbf{R4: }Facilitate cross-community collaborations  & 0 & 0 & $++$ & 0 & 0 & 0 & 0  \\  \midrule 
\textbf{D1: }Built to exist in an ecosystem  & 0 & 0 & 0 & 0 & 0 & 0 & 0  \\ 
\textbf{D2: }Leverage open-source projects  & $+$ & $+$ & 0 & $+$ & $+$ & $+$ & $+$  \\ \bottomrule 
\end{tabular}\vspace{-5pt}
\end{center}
\end{table*}
\normalsize

Over the past decade, the self-adaptive systems community has developed a substantial number of artifacts covering a wide variety of purposes and domains.\footnote{\url{http://www.self-adaptive.org/exemplars/}} 
In a recent community effort, we identified a set of guidelines for industry-relevant artifacts to perform and evaluate research in self-adaptation~\cite{10.1145/3561846.3561852}. These guidelines, which  leverage the empirically-grounded results of the survey on the use of self-adaptation in industry~\cite{3524844.3528077,survey-report}, are divided into three groups: required, recommended, and desirable guidelines. 
Our interest is to use these guidelines to assess the industrial relevance of a sample of existing artifacts and hence the maturity of research results for industrial adoption obtained by using the artifacts.  
We then take a position on whether or not solutions developed by the self-adaptive systems community are mature enough for industrial adoption. We conclude with a reflection on our position.

\section{Industrial Relevance of SEAMS Artifacts}
We evaluated the industry relevance of the seven most recent artifacts that were presented at SEAMS 2021 and 2022: 
Platooning LEGOs (PLEGOs)~\cite{9462025} focuses on platooning autonomous cars implemented with LEGO,   
Body Sensor Network (BSN)~\cite{9462020} targets monitoring the health status of patients, 
RoboMAX~\cite{RoboMax} focuses on exemplars for robotic missions, 
RDMSim~\cite{9462042} targets decision-making for Remote Data Mirroring systems,
Emergent Web Server (EWS)~\cite{9799868} focuses on learning optimal architectural compositions, 
SEAByTE~\cite{9799807} focuses on automating A/B testing pipelines of a micro-service based system, 
and Simdex~\cite{9800016} targets an adaptive backend that dispatches computing jobs among multiple workers.

Table~\ref{tab:overview} shows an overview of the industry-relevance of the artifacts using the guidelines~of~\cite{10.1145/3561846.3561852}. 
The data was collected from the artifact papers and documentation by two researchers and in the case of differences in opinions discussed until a consensus was reached. 
The score $++$ means that an artifact explicitly deals with a guideline; $+$ means that an artifact either implicitly or only partially deals with a guideline, and the score $0$  means that the guideline is not considered.

\section{Position Statement and Reflection}

From the required guidelines, we observe that only ''Offer industry-relevant metrics'' is partially supported by the selected artifacts and ''Use an open policy'' is generally supported. The other required guidelines are in general not well supported. Similarly, for the recommended guidelines, we observe that ''Integrate with standard technologies'' is mostly partially supported, and for the desirable guidelines only ''Leverage open-source projects'' is partially supported. This brings us to the conclusion that the industrial relevance of a sample of existing artifacts is limited and hence the maturity of research results obtained by using them in terms of industrial adoption is expected to be rather weak. Consequently, our answer to the question of whether the solutions developed by the self-adaptive systems community \mbox{are mature enough for industrial adoption is: \textit{not yet!}}

It is important to put our position in perspective. Since we only used a sample of artifacts, we need to take care of generalization (external validity). Further, our position takes a particular angle on the question, namely the industrial relevance of existing artifacts. We connect this relevance to the maturity of research results obtained by using the artifacts for industrial adoption (construct validity). Also, research results may be evaluated by other means as artifacts, and the guidelines for industrial relevance may not accurately reflect or imply the maturity of solutions for industrial adoption. 

It is also important to highlight that the existing artifacts were not developed with the explicit aim to be industry-relevant. Nevertheless, since our community actively promotes the use of artifacts for evaluating research results, we believe that the insights obtained from our limited study and hence our position is important for the future of our community and can contribute to the debate regarding the maturity of research results on self-adaptation for industrial adoption.   

\section{Conclusion}
The evaluation of a sample of artifacts produced by our community based on a set of empirically-grounded guidelines indicates that their industrial relevance and hence the maturity for industrial adoption of the research results obtained by using the artifacts is limited. We advice future artifact developers to take into account: (1) relevant surrounding elements (real-world data, humans-on-the-loop, industrial adoption), (2) industrial scale, and (3) industry-relevant metrics. Additionally, artifact developers may consider process-related artifacts, generality and extensibility, support for cross-community collaboration, and integration into existing ecosystems.

\bibliographystyle{IEEEtran}
\bibliography{main}

\end{document}